\documentclass{Interspeech2024}
\usepackage{amsmath,amsfonts}
\usepackage[ruled,vlined]{algorithm2e}
\usepackage{array}
\usepackage{float}
\usepackage{textcase}
\usepackage[caption=false,font=normalsize,labelfont=sf,textfont=sf]{subfig}
\usepackage{textcomp}
\usepackage{stfloats}
\usepackage{url}
\usepackage{verbatim}
\usepackage{graphicx}
\usepackage{cite}
\usepackage{makecell}
\usepackage{booktabs}
\usepackage{multirow,booktabs,pifont}
\newcommand{\xmark}{\ding{55}}
\newcommand{\cmark}{\ding{51}}
\usepackage{hyperref} 
\usepackage{diagbox}
\usepackage{colortbl}
\usepackage{xcolor}
\newcommand{\tabincell}[2]{\begin{tabular}
{@{}#1@{}}#2\end{tabular}}
\newcommand{\deepgreen}[1]{\textcolor{green!70!black}{#1}}
\newcommand{\brownred}[1]{\textcolor{brown!90!red}{#1}}
\definecolor{lightred}{RGB}{255,20,0}

\makeatletter

\newcommand{\Rmnum}[1]{\expandafter\@slowromancap\romannumeral #1@}
\makeatother

\interspeechcameraready

\title{Joint Speaker Features Learning for Audio-visual Multichannel Speech Separation and Recognition}

\name[affiliation={1}]{Guinan}{Li}
\name[affiliation={1}]{Jiajun}{Deng}
\name[affiliation={1}]{Youjun}{Chen}
\name[affiliation={2}]{Mengzhe}{Geng}
\name[affiliation={1}]{Shujie}{Hu}
\name[affiliation={3}]{Zhe}{Li}
\name[affiliation={1}]{Zengrui}{Jin}
\name[affiliation={1}]{Tianzi}{Wang}
\name[affiliation={4}]{Xurong}{Xie}
\name[affiliation={1}]{Helen}{Meng}
\name[affiliation={1}]{Xunying}{Liu}

\address{
  $^1$The Chinese University of Hong Kong; $^2$National Research Council Canada; $^3$The Hong Kong Polytechnic University; $^4$Institute of Software, Chinese Academy of Sciences}
\email{\{gnli,jjdeng, yjchen, sjhu, zrjin, twang, hmmeng, xyliu\}@se.cuhk.edu.hk, Mengzhe.Geng@nrc-cnrc.gc.ca, lizhe.li@connect.polyu.hk,  xurong@iscas.ac.cn}


\begin{document}

\maketitle

\begin{abstract}
   This paper proposes joint speaker feature learning methods for zero-shot adaptation of audio-visual multichannel speech separation and recognition systems. xVector and ECAPA-TDNN speaker encoders are connected using purpose-built fusion blocks and tightly integrated with the complete system training. Experiments conducted on LRS3-TED data simulated multichannel overlapped speech suggest that joint speaker feature learning consistently improves speech separation and recognition performance over the baselines without joint speaker feature estimation. Further analyses reveal performance improvements are strongly correlated with increased inter-speaker discrimination measured using cosine similarity. The best-performing joint speaker feature learning adapted system outperformed the baseline fine-tuned WavLM model by statistically significant WER reductions of 21.6\% and 25.3\% absolute (67.5\% and 83.5\% relative) on Dev and Test sets after incorporating WavLM features and video modality. 
\end{abstract}
 
\section{Introduction}
Despite the rapid progress of automatic speech recognition (ASR) in recent decades, accurate recognition of cocktail party speech 
\cite{qian2018past} remains a highly challenging task to date. To this end, microphone arrays play a key role in state-of-the-art speech separation and recognition systems designed for such data \cite{haeb2020far}. The required array beamforming techniques used to perform multichannel signal integration are implemented as time or frequency domain filters. These are represented by time-domain delay and sum \cite{anguera2007acoustic, xu2020neural}, frequency-domain minimum variance distortionless response (MVDR) \cite{souden2009optimal} and generalized eigenvalue (GEV) \cite{warsitz2007blind} based multichannel integration approaches.

In recent years, end-to-end DNN-based microphone array beamforming techniques represented by a) neural time-frequency (TF) masking approaches \cite{bahmaninezhad2019}; b) neural Filter and Sum methods \cite{sainath2017multichannel,xiao2016deep}; and c) mask-based MVDR  \cite{yoshioka2018multi} and generalized eigenvalues (GEV) \cite{heymann2017beamnet} approaches have been widely adopted. In addition, incorporating visual information into either multi speech separation front-ends alone \cite{gu2020multi}, or further into speech recognition back-ends \cite{li2023audio}, can further improve the overall system performance.

Natural speech represented by cocktail party speech is highly heterogeneous. To this end, speaker adaptation techniques have been widely studied as powerful solutions to customise ASR systems for individual users. These include, but not limited to: 1) auxiliary speaker embedding based approaches \cite{saon2013speaker, snyder2018x, vesely2016sequence, kanda2021end, kanda2021investigation}, e.g. iVector \cite{saon2013speaker} and xVector \cite{snyder2018x}; 2) feature transformation based methods, e.g., feature-space MLLR \cite{gales1998maximum}; and 3) model-based methods \cite{ochiai2018speaker, swietojanski2016learning, ochiai2014speaker, deng2023confidence} that estimate speaker dependent (SD) adapter parameters implemented as, e.g. learning hidden unit contributions (LHUC) \cite{swietojanski2016learning}, during speaker adaptive training (SAT) and test-time unsupervised adaptation \cite{ochiai2014speaker, deng2023confidence}. 

Recent researches have also demonstrated that overlapping speech separation \cite{vzmolikova2019speakerbeam, ochiai2019multimodal, Wang2019voicefilter,ju2022tea, ju2023tea,xu2020spex,ge2020spex+,eskimez2022personalized, taherian2022one} and recognition \cite{vzmolikova2019speakerbeam, Wang2019voicefilter, eskimez2022personalized, taherian2022one} systems also benefit from modelling speaker-level features . These include, but not limited to, SpeakerBeam \cite{vzmolikova2019speakerbeam, ochiai2019multimodal}, VoiceFilter  \cite{Wang2019voicefilter}, TEA-PSE \cite{ju2022tea, ju2023tea}, SpEx \cite{xu2020spex,ge2020spex+} and PSE \cite{eskimez2022personalized, taherian2022one}. However, these prior studies suffer from several limitations: \textbf{1)} Lack of tight integration between speaker feature learning and backbone speech enhancement-recognition model training. Speaker features are learned either separately and de-coupled from the backbone speech enhancement-recognition model without any form of joint training \cite{Wang2019voicefilter, ju2022tea, ju2023tea, eskimez2022personalized, taherian2022one}, or only partially jointly learned with the speech enhancement front-end alone, while separated from the speech recognition back-end \cite{vzmolikova2019speakerbeam, ochiai2019multimodal, xu2020spex,ge2020spex+}. 
\textbf{2)} Lack of zero-shot, instantaneous adaptation methods for speech enhancement models. Instead the commonly used enrollment-based speaker adaptation approaches \cite{vzmolikova2019speakerbeam, ochiai2019multimodal, Wang2019voicefilter,ju2022tea, ju2023tea,xu2020spex,ge2020spex+,eskimez2022personalized, taherian2022one} require clean speech samples to be explicitly recorded at the onset of user personalization. This not only incurs processing latency but also leads to privacy concerns. 
\textbf{3)} The efficacy of speaker adaptation was predominantly evaluated on speech separation systems alone \cite{vzmolikova2019speakerbeam, ochiai2019multimodal, Wang2019voicefilter,ju2022tea, ju2023tea,xu2020spex,ge2020spex+,eskimez2022personalized, taherian2022one}, while there is a lack of holistic incorporation of the speaker features into both the speech separation front-end and recognition back-end components. 

To this end, this paper proposes joint speaker features learning approaches for zero-shot,  instantaneous adaptation of audio-visual multichannel speech separation and recognition systems. xVector \cite{snyder2018x} or ECAPA-TDNN \cite{desplanques20_interspeech} speaker feature extraction modules are connected with the backbone system using purpose-built fusion blocks, and tightly integrated with the complete system training. The resulting systems support zero-shot, enrollment-free and on-the-fly adaptation to unseen speakers without requiring pre-recorded user data. Experiments conducted on simulated multichannel overlapped and noisy speech data constructed using the benchmark LRS3-TED \cite{afouras2018lrs3}  dataset suggest that joint speaker features learning consistently produces speech enhancement and recognition performance improvements over the baselines without joint speaker feature estimation.
Further analyses reveal that these performance improvements are strongly correlated with the increase of inter-speaker discrimination measured using cosine similarity.  
The best-performing joint speaker feature learning adapted system outperformed the baseline fine-tuned WavLM model by statistically significant WER reductions of \textbf{21.6\%} and \textbf{25.3\% absolute} (\textbf{67.5\%} and \textbf{83.5\% relative}) on Dev and Test sets after incorporating WavLM features and video modality.
The main contributions of this paper are summarized below: 

\textbf{1)} To the best of our knowledge, this paper presents the first use of joint speaker features learning approaches for zero-shot user adaptation of audio-visual multichannel speech separation and recognition systems. In contrast, in prior researches speaker features are learned either separately and de-coupled from the backbone system without any form of joint training \cite{Wang2019voicefilter, ju2022tea, ju2023tea, eskimez2022personalized, taherian2022one}, or only partially jointly learned with the speech enhancement front-end alone, while separated from the speech recognition backend \cite{vzmolikova2019speakerbeam, ochiai2019multimodal, xu2020spex,ge2020spex+}. 

\textbf{2)}  To the best of our knowledge, this paper pioneers the use of zero-shot, enrollment-free speaker adaptation for speech separation and recognition tasks. In contrast, the related prior studies only focus on less practical enrollment-based adaptation techniques \cite{vzmolikova2019speakerbeam, ochiai2019multimodal, Wang2019voicefilter,ju2022tea, ju2023tea,xu2020spex,ge2020spex+,eskimez2022personalized, taherian2022one} that require clean speech samples to be explicitly recorded at the onset of user personalization.

 \textbf{3)} This paper presents the first investigation of the complete incorporation of speaker features into all the components of a complete end-to-end audio-visual multichannel speech separation and recognition system. In contrast, prior researches consider speaker adaptation of either the speech separation front-end \cite{vzmolikova2019speakerbeam, ochiai2019multimodal, Wang2019voicefilter,ju2022tea, ju2023tea,xu2020spex,ge2020spex+,eskimez2022personalized, taherian2022one, gu2020multi} alone, or the speech recognition back-end \cite{saon2013speaker, vesely2016sequence, kanda2021end, kanda2021investigation, gales1998maximum, swietojanski2016learning, ochiai2014speaker,  deng2023confidence} only.

\begin{figure*}[htbp]
\vspace{-0.5cm}
    \centering
    \setlength{\abovecaptionskip}{0pt plus 1pt minus 3pt}
    \includegraphics[scale=0.45]{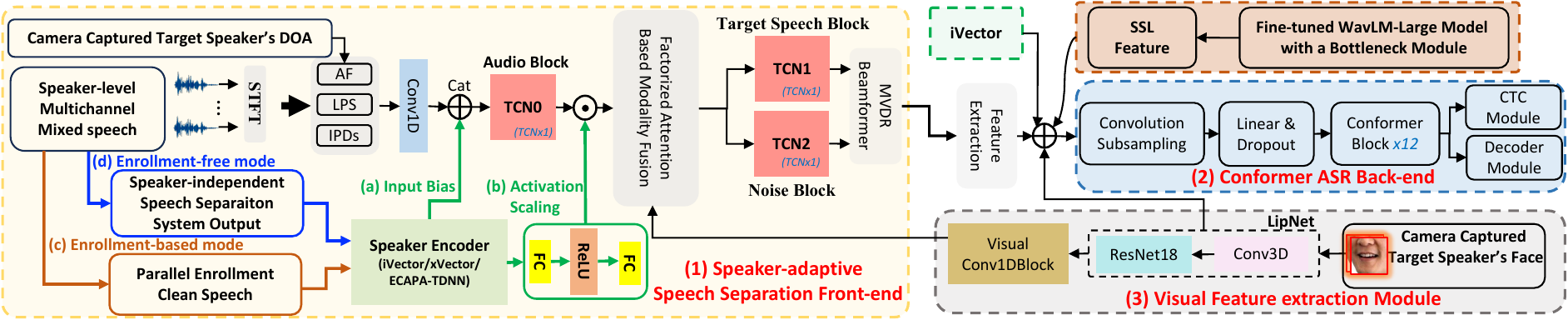}
    \caption{Example of joint speaker features learning for an audio-visual multichannel speech separation and recognition system including the following components: \textbf{\textcolor{red}{(1) Speaker-adaptive speech separation front-end}} implemented using TCNs and mask-based MVDR beamforming; \textbf{\textcolor{red}{(2) Conformer ASR Back-end}}; and \textbf{\textcolor{red}{(3) Visual Feature extraction Module}}. The Speaker Encoder (iVector, xVector or ECAPA-TDNN) module is connected with the backbone system using purpose-built fusion blocks based on either \textbf{\deepgreen{(a) Input Bias}} or \textbf{\deepgreen{(b) Activation Scaling}}, and tightly integrated with complete system training. Speaker adaptation is performed in either \textbf{\brownred{(c) enrollment-based mode}} \cite{vzmolikova2019speakerbeam, ochiai2019multimodal, Wang2019voicefilter,ju2022tea, ju2023tea,xu2020spex,ge2020spex+,eskimez2022personalized, taherian2022one} requiring pre-recorded speaker-level parallel clean-noisy speech; or \textbf{\textcolor{blue}{(d) zero-short, enrollment-free mode}} that does not require pre-recorded speaker-level clean speech data. ``Cat” and ``$\odot$” denote the concatenation and element-wise product operation, respectively.
    }
    \label{fig:sep_asr_adapt}
 \vspace{-0.7cm}
\end{figure*}

\vspace{-0.2cm}
\section{Audio-visual Speech Separation}
\vspace{-0.1cm}
\subsection{Mask-based MVDR}
\vspace{-0.1cm}
In MVDR beamforming \cite{souden2009optimal, warsitz2007blind}, 
a linear filter $\mathbf{w}(f) \in \mathbb{C}^{R}$ is applied to the multichannel mixture speech short-time Fourier transform (STFT) spectrum $\mathbf{y}(t, f) \in \mathbb{C}^{R}$ to produce the filtered output ${\hat S}(t,f)$ as:
\vspace{-0.2cm}
\begin{equation}
\!{{\small \hat{S}(t,\!f)} \!= {\mathbf{w}(f)^H\mathbf{y}(t,\!f)} \!=\! {\underbrace{\mathbf{w}(f)^H \mathbf{x}(t, \!f)}_{\text{target speech component}}}\!\!\!+\!{\underbrace{\mathbf{w}(f)^H\mathbf{n}(t,\!f)}_{\text{residual noise}}}}\!,
 \label{equation:linear_filtering_2}
\end{equation} 
where $R$ denotes the channel number of a microphone array.  $t$ and $f$ denote the indices of time and frequency bins, respectively.  $(\cdot)^{H}$ denotes the conjugate transpose operator. $\mathbf{x}(t, f) \in \mathbb{C}^{R}$ is a complex vector containing the clean speech signals. $\mathbf{n}(t, f) \in \mathbb{C}^{R}$ represents either the interfering speaker’s speech or additive background noise alone, or a combination of both. 

By minimizing the residual noise output while imposing a distortionless constraint on the target speech \cite{souden2009optimal}, 
the MVDR beamforming filter is estimated as 
\vspace{-0.2cm}
\begin{align}
{\small \mathbf{w}(f) 
= \frac{\boldsymbol{\Phi}_n(f)^{-1} \boldsymbol{\Phi}_x(f)} {\operatorname{tr}\left(\boldsymbol{\Phi}_n(f)^{-1} \boldsymbol{\Phi}_x(f)\right)} \mathbf{u}_{r}},\label{equation:mvdr_filter_2}
\end{align}
where $\mathbf{u}_r=[0,\ldots, 1, \ldots, 0]^T \in \mathbb{R}^{R}$ is a one-hot reference vector where its $r$-th component equals to one. $\operatorname{tr}(\cdot)$ denotes the trace operator.
Without loss of generality, we select the first channel as reference in this paper.
The target speaker power spectral density (PSD) matrix $\boldsymbol{\Phi}_x(f)$  and noise-specific PSD matrix $\boldsymbol{\Phi}_n(f)^{-1}$ are computed using DNN predicted complex TF masks. More details can be found in the paper \cite{li2023audio}.
\vspace{-0.2cm}
\subsection{Audio and Visual Modality Fusion}
\vspace{-0.1cm}
\label{subsecton:AV_modality}
\noindent\textbf{Audio modality:}
Three types of audio features including the complex STFT spectrum of all the microphone array channels, the inter-microphone phase differences (IPDs) \cite{yoshioka2018recognizing} and angle feature (AF) \cite{chen2018multi} are adopted as the audio inputs (Fig. \ref{fig:sep_asr_adapt}, upper left, in light grey).  Following 
\cite{li2023audio}, temporal convolutional networks (TCNs) \cite{luo2019conv} are used. Each TCN block is stacked by 8 Dilated1DConvBlock with exponentially increased dilation factors $2^0, 2^1, \ldots ., 2^7$. The log-power spectrum (LPS) features of the reference microphone channel are concatenated with the IPDs and AF features before being fed into a Conv1D module followed by a single TCN-based Audio Block to compute audio embeddings.

\noindent\textbf{Visual modality:}
The target speaker's lip region obtained via face tracking is fed into a LipNet \cite{afouras2018deeplip} containing a 3D convolution layer (Fig. \ref{fig:sep_asr_adapt}, bottom right, in pink) and 18-layer ResNet \cite{he2016deep} (Fig. \ref{fig:sep_asr_adapt}, bottom right, in light turquoise) to extract visual features before being fed into a Visual Conv1DBlock. Then, the output of the Visual Conv1DBlock is up-sampled and time synchronised with the audio frames via linear interpolation to compute visual embeddings.

\noindent\textbf{Modality fusion:}
A factorized attention-based modality fusion module (Fig. \ref{fig:sep_asr_adapt}, middle, in grep) following the prior works in \cite{li2023audio} is utilized to integrate the audio and visual embeddings. 

\noindent\textbf{Separation Network Training Cost Function:} 
The mask-MVDR based multichannel speech separation network is trained separately to maximize the SISNR metric before joint fine-tuning using the back-end ASR error loss in Eqn. (\ref{eqn:cfm-hybrid-loss}).

\vspace{-0.2cm}
\section{Speaker Adaptation of Audio-visual Speech Separation Front-end}
\vspace{-0.1cm}
Speaker features based on either iVector  \cite{saon2013speaker}, xVector \cite{snyder2018x} or ECAPA-TDNN \cite{desplanques20_interspeech} are connected with the speech separation backbone system using 
either input bias or activation scaling. Among these, xVector and ECAPA-TDNN speaker encoders are tightly integrated with the entire speech enhancement-recognition model training. Speaker adaptation is performed in either enrollment-based or zero-shot enrollment-free modes.
\vspace{-0.6cm}
\subsection{Speaker Feature Fusion}
The utterance-level speaker features extracted from each speaker encoder are incorporated into the backbone system using either \textbf{input bias} or \textbf{activation scaling} with purpose-built fusion blocks for adaptation. \textbf{a) Input bias fusion} refers to the utterance-level speaker features being concatenated directly with the output audio embeddings from the Conv1D module along the feature dimension (Fig. \ref{fig:sep_asr_adapt}, middle, green line marked with "\deepgreen{Input Bias}"). \textbf{b) Activation scaling fusion} feeds the speaker features through a fusion module before being further applied to the TCN outputs using element-wise product. The fusion module consists of a fully connected (FC) layer with 256 x 200 dimensions and a ReLU activation followed by an output FC layer with 256 output units (Fig.~1, middle, green line marked with "\deepgreen{Activation Scaling}”).

\vspace{-0.3cm}
\subsection{Speaker Feature Learning}
\vspace{-0.1cm}
Speaker features are learned either \textbf{separately from} or \textbf{tightly integrated with} the backbone speech enhancement-recognition model training. \textbf{a)} In \textbf{non-joint speaker feature learning}, speaker encoders are trained in an offline manner using the speaker recognition error loss\cite{Wang2019voicefilter, ju2022tea, ju2023tea, eskimez2022personalized, taherian2022one}. The features extracted from these speaker encoders are fed into the backbone speech enhancement-recognition system using the above fusion methods. \textbf{b)} In \textbf{joint speaker feature learning}, xVector or ECAPA-TDNN speaker encoders are tightly integrated with the backbone enhancement-recognition model. They are updated in turn using the SI-SNR cost for speech separation front-end alone, and followed by ASR cost for the entire enhancement-recognition system. Further analyses in Sec. 5.2 reveal that these performance improvements brought by jointly speaker feature learning are strongly correlated with the increase of inter-speaker discrimination measured using cosine similarity. 

 \vspace{-0.3cm}
\subsection{Adaptation Supervision}
\vspace{-0.1cm}
Speaker adaptation is performed in either \textbf{enrollment-based} mode [25–33] or \textbf{zero-short, enrollment-free} mode. \textbf{a)} In \textbf{enrollment-based supervised} adaptation \cite{vzmolikova2019speakerbeam, ochiai2019multimodal, Wang2019voicefilter,ju2022tea, ju2023tea,xu2020spex,ge2020spex+,eskimez2022personalized, taherian2022one} (Fig.~\ref{fig:sep_asr_adapt}, bottom left, \brownred{light brown box}), pre-recorded speaker-level parallel clean-noisy speech is required at the onset of user personalization. For example, SpeakeBeam\cite{vzmolikova2019speakerbeam} uses 100\% of the test set data' ground truth clean speech to construct speaker-level enrollment data\footnote{https://github.com/BUTSpeechFIT/speakerbeam}. \textbf{b)} In \textbf{enrollment-free, zero-shot} adaptation (Fig.~\ref{fig:sep_asr_adapt}, middle left, \textcolor{blue}{blue box}), speaker-level parallel clean-noisy speech is not required. The target clean speech is replaced by the enhanced speech outputs produced by a speaker-independent speech separation system before being fed into the speaker feature encoder. This form of zero-shot speaker feature learning directly from untranscribed overlapped and noisy speech alone alleviates the processing latency in speaker clean-noisy parallel speech pre-recording and privacy issues brought by enrollment-based adaptation.

\vspace{-0.3cm}
\section{Audio-visual Conformer ASR Back-end}
\vspace{-0.1cm}
Speech separation  outputs are used to extract
 Mel-filterbank features.  They are concatenated with visual features and fed into the ASR back-end. The hybrid CTC-attention Conformer ASR model \cite{gulati2020conformer} consists of an encoder module and a decoder module.
 Fig. \ref{fig:sep_asr_adapt} (middle right, in light blue) shows an example of a Conformer ASR system. More details can be found in \cite{gulati2020conformer}. 
The following multi-task criterion interpolation between the CTC and attention error costs \cite{watanabe2017hybrid} is used in Conformer model training, 
\vspace{-0.3cm}
\begin{equation} \label{eqn:cfm-hybrid-loss}
\mathcal{L_\text{ASR}}=(1-\beta) \mathcal{L}_{a t t} + \beta \mathcal{L}_{c t c},
\end{equation}
where $\beta \in[0,1]$ is a tunable hyper-parameter and empirically set as $0.3$ for training and $0.4$ for recognition in this paper. \\
\noindent
\textbf{End-to-end joint fine-tuning of speech separation and the recognition} components \cite{li2023audio} using the ASR cost function of Eqn. \ref{eqn:cfm-hybrid-loss} is applied in this paper.\\
\noindent\textbf{iVector-based Speaker Adaptation:}
iVector features extracted with a 100ms  window are up-sampled to be time synchronised with the  Mel-filterbank audio frames \cite{geng2022speaker} (Fig.~\ref{fig:sep_asr_adapt}, top middle, in light green). Then, the iVector, audio and visual features are concatenated together and fed into the ASR back-end.\\
\noindent\textbf{SSL pre-trained WavLM features:}
The enhanced speech fine-tuned WavLM-Large models \cite{chen2022wavlm} contain a Bottleneck Module \cite{hu2023exploring}. It is used to extract SSL pre-trained features that are fed into Conformer ASR systems via input feature concatenation with Mel-filterbanks (Fig.~\ref{fig:sep_asr_adapt}, top left, in light orange).
\vspace{-0.3cm}

\begin{table*}[h!]
\vspace{-0.7cm}
\centering
\caption{Performance of joint speaker features learning for audio-only speech separation and recognition systems in either enrollment-based or enrollment-free mode. The ASR back-end used to evaluate the WER metric is fine-tuned in a pipelined manner by the speech separation output.
``Enroll.", ``Spk. Feat.", and ``O.V." denote  ``enrollment",  ``speaker feature" and ``overall", respectively. 
``$\dagger$" represents a statistically significant (MAPSSWE, $\alpha$=0.05\cite{Gillick1989SomeSI} ) WER difference over the SI baseline (sys. 1).\\
}
\vspace{-0.6cm}
\label{tab:table2}
\resizebox{1.89\columnwidth}{!}{
\begin{tabular}{c|c|c|c|c|c|c|c|c|c|c} 
 \toprule[1.2pt]
 \multirow{3}{*}{Sys.}&   \multicolumn{4}{c|}{Speaker Adaptation for Speech Separation}  & \multicolumn{6}{c}{SISNR$(\uparrow)$/PESQ$(\uparrow)$/STOI$(\uparrow)$/WER $(\downarrow)$} \\
 \cline{2-11}
 &   \multirowcell{2}{\scriptsize{Joint Spk.}\\ \scriptsize{Feat. Learning}} & \multirowcell{2}{\scriptsize{Adaptation}\\ \scriptsize{Supervision}} &  \multirowcell{2}{\scriptsize{Spk. Feat.} \\ \scriptsize{Fusion} }  & \multirowcell{2}{\scriptsize{Spk. Feat.} \\ \scriptsize{Encoding}}   & \multicolumn{5}{c|}{Dev} &  Test \\
 \cline{6-11}
&   & & &  & \multirowcell{1}{\scriptsize [$0^{\circ}$, $15^{\circ}$)}  & \multirowcell{1}{\scriptsize [$15^{\circ}$, $45^{\circ}$)} &\multirowcell{1}{\scriptsize [$45^{\circ}$, $90^{\circ}$) }& \multirowcell{1}{\scriptsize [$90^{\circ}$, $180^{\circ}$)}  & \multirowcell{1}{\scriptsize {O.V.}} & \multirowcell{1}{\scriptsize {O.V.}}  \\
\midrule[1pt]
\multicolumn{5}{c|}{\multirowcell{1}{The raw first channel of overlapped speech}} & 0.05/1.92/65.16/55.4  &  -0.05/1.91/64.99/56.3 & -0.02/1.87/65.25/53.7  &-0.04/1.95/65.34/55.0  &  -0.02/1.91/65.18/55.1 & 0.06/1.83/64.52/48.6 \\
\midrule[1pt]
1 &  \multicolumn{4}{c|}{\multirowcell{1}{Speaker-independent (SI) baseline}} & 4.26/2.44/76.37/48.8  & 9.16/2.86/85.45/29.6 &10.58/2.95/87.79/26.0 & 10.84/3.05/88.80/24.7 &  8.69/2.82/84.56/32.5 & 7.62/2.75/84.94/25.1\\
\midrule[1pt]
2  & \xmark  & \multirowcell{12}{Enroll.}  &  \multirowcell{5}{Input \\ Bias } &  \multirowcell{1}{\scriptsize{iVector}}   & 6.56/\textbf{2.60/80.80/41.2}  & \textbf{9.71/2.89}/85.89/\textbf{29.0} & \textbf{10.91/2.97}/88.05/\textbf{25.4} & \textbf{11.13/3.06}/88.94/24.6  &  \textbf{9.56/2.88/85.90}/30.2$^\dagger$ & \textbf{8.72/2.83/87.08/21.6}$^\dagger$\\ 
3  & \xmark &   & &  \multirowcell{1}{\scriptsize{xVector}} &  \textbf{6.63}/2.59/80.66/42.0 & 9.43/2.88/85.79/29.4 & 10.61/2.95/87.79/25.9 & 10.93/3.04/88.80/24.5 & 9.39/2.86/85.74/30.6$^\dagger$ &  8.50/2.81/86.79/22.6$^\dagger$ \\
4 &  \xmark & &  & \multirowcell{1}{\scriptsize{ECAPA-TDNN}} & 5.73/2.53/79.19/44.1 & 9.35/2.87/85.63/\textbf{29.0} & 10.70/2.96/87.93/26.3  & 10.98/3.05/88.86/24.3  & 9.18/2.85/85.38/31.1$^\dagger$   & 8.30/2.79/86.48/23.0$^\dagger$ \\
5 & \cmark &  &  & \multirowcell{1}{\scriptsize{xVector}}  & 6.37/2.58/80.17/42.9 & 9.56/\textbf{2.89/85.97}/29.1 & 10.85/\textbf{2.97/88.10/25.4}& 11.07/\textbf{3.06/88.97/24.2} & 9.45/2.87/85.78/\textbf{29.7}$^\dagger$ & 8.56/2.82/86.84/21.9$^\dagger$ \\
6 & \cmark &   &   &  \multirowcell{1}{\scriptsize{ECAPA-TDNN}} & 6.04/2.55/79.54/43.5 & 9.38/2.87/85.73/29.7 & 10.78/2.96/88.03/26.0& 11.07/\textbf{3.06/88.97}/24.3  & 9.30/2.86/85.54/31.0$^\dagger$&  8.46/2.80/86.78/22.7$^\dagger$\\
\cline{1-2}
\cline{4-11}
7 &\xmark  & &  \multirowcell{5}{Act. \\ Scaling} &  \multirowcell{1}{\scriptsize{iVector}}   & \textbf{7.13/2.63/81.51/40.4} & 9.34/2.87/85.54/30.3 & 10.67/2.95/87.67/26.3 & 10.86/3.04/88.70/25.0 & \textbf{9.49/2.87/85.84}/30.6$^\dagger$ & \textbf{8.74/2.82/87.17/22.4}$^\dagger$\\ 
8 &\xmark &  &  &  \multirowcell{1}{\scriptsize{xVector}} &  6.59/2.58/80.37/42.4 & 9.31/2.87/85.62/29.6 & 10.63/\textbf{2.96}/87.95/26.6 & 10.99/\textbf{3.05}/88.87/24.6 & 9.37/2.86/85.68/30.9$^\dagger$ &  8.47/2.80/86.79/23.0$^\dagger$\\
9 & \xmark &  &  & \multirowcell{1}{\scriptsize{ECAPA-TDNN}} &  6.58/2.58/80.73/43.1 & 9.35/2.87/85.63/29.7 & 10.68/2.95/87.91/26.5 & \textbf{11.02}/3.04/88.83/24.4 & 9.40/2.86/85.75/31.1$^\dagger$ & 8.47/2.80/86.71/22.7$^\dagger$ \\
10 & \cmark  & & &  \multirowcell{1}{\scriptsize{xVector}} & 6.36/2.57/80.20/41.7 & 9.47/\textbf{2.88}/85.78/\textbf{29.0} & \textbf{10.74/2.96/87.97}/26.0 & 10.98/\textbf{3.05/88.93}/24.3 & 9.38/\textbf{2.87}/85.70/\textbf{30.4}$^\dagger$  & 8.55/2.81/86.87/22.7$^\dagger$\\
11 & \cmark  & &   & \multirowcell{1}{\scriptsize{ECAPA-TDNN}} & 6.62/2.58/80.67/42.2  & \textbf{9.54/2.88/85.89}/29.3  & 10.67/\textbf{2.96}/87.89/\textbf{25.8}  &  10.99/\textbf{3.05}/88.89/\textbf{24.1} &  9.44/\textbf{2.87}/85.82/30.5$^\dagger$ & 8.57/2.81/86.82/22.6$^\dagger$ \\
\cline{1-11}
12 &  \xmark &  \multirowcell{12}{Enroll. \\ Free}  &  \multirowcell{5}{Input \\ Bias} &  \multirowcell{1}{\scriptsize{iVector}}  
 &\textbf{6.01/2.57/80.09/42.0} & 9.40/2.89/85.72/29.0 & 10.80/2.96/88.05/\textbf{25.2} & 10.95/3.06/88.91/24.6 & \textbf{9.28/2.87/85.67/30.3}$^\dagger$& 7.97/2.78/85.84/23.6$^\dagger$\\ 
13  &  \xmark  &  &  &  \multirowcell{1}{\scriptsize{xVector}} & 4.82/2.50/77.21/45.4 & 9.65/\textbf{2.90}/86.00/28.8 & 10.82/2.97/88.04/25.6 & 11.05/3.06/89.00/\textbf{24.2} & 9.06/2.86/85.03/31.2$^\dagger$& 8.16/2.80/85.87/23.7$^\dagger$\\
14 & \xmark  & &   & \multirowcell{1}{\scriptsize{ECAPA-TDNN}} &  4.94/2.51/77.41/45.6 & 9.55/2.89/85.82/29.4 & 10.81/2.96/88.01/26.7 & 11.04/3.05/88.90/24.4 & 9.07/2.85/85.00/31.7$^\dagger$  & 8.20/2.80/85.92/23.2$^\dagger$\\
15 & \cmark &  & & \multirowcell{1}{\scriptsize{xVector}} &   4.98/2.51/77.47/44.1 & 9.68/2.89/85.93/28.8 & 10.95/2.97/88.06/25.6 & 11.16/3.06/89.02/24.7 & 9.17/2.86/85.09/31.0$^\dagger$ & 8.25/2.80/85.84/23.2$^\dagger$\\ 
16 &\cmark   &   &  & \multirowcell{1}{\scriptsize{ECAPA-TDNN}}  &   5.12/2.52/77.56/44.9  & \textbf{9.80/2.90/86.18/28.4}  & \textbf{10.97/2.98/88.19}/25.8  & \textbf{11.24/3.07/89.09}/24.3 & 9.26/2.86/85.22/31.1$^\dagger$  & \textbf{8.31/2.81/86.13/22.7}$^\dagger$ \\
\cline{1-2}
\cline{4-11}
17 & \xmark & &  \multirowcell{5}{Act. \\ Scaling} &  \multirowcell{1}{\scriptsize{iVector}}  &   \textbf{6.45/2.59/80.28/42.5} & 9.15/2.86/85.21/31.0 & 10.60/2.95/87.63/26.2  & 10.81/3.04/88.64/24.9 & 9.24/2.86/\textbf{85.42}/31.3$^\dagger$ &  8.15/2.78/85.89/23.6$^\dagger$\\ 
18 & \xmark &  &  &  \multirowcell{1}{\scriptsize{xVector}} &  4.79/2.49/77.09/46.3 & 9.42/2.88/85.68/30.2 & 10.61/2.96/87.89/25.9 & 11.00/3.05/88.89/24.8  & 8.94/2.85/84.85/32.0 & 8.03/2.78/85.47/24.2 $^\dagger$  \\
19 & \xmark  &  &  & \multirowcell{1}{\scriptsize{ECAPA-TDNN}}  & 4.79/2.50/77.09/45.7 & 9.38/2.88/85.57/30.6 & 10.80/2.97/88.00/\textbf{25.8} & 11.07/\textbf{3.07}/89.00/\textbf{24.4} & 8.99/2.85/84.88/31.8$^\dagger$  & 8.08/2.79/85.65/23.6$^\dagger$\\
20 &  \cmark  & & &   \multirowcell{1}{\scriptsize{xVector}}  &  5.08/2.52/77.51/44.5 & 9.80/2.90/\textbf{86.14}/28.9 &  \textbf{10.96/2.98}/88.21/\textbf{25.8} & 11.22/3.06/89.06/24.9 & 9.24/2.86/85.19/31.2$^\dagger$& 8.29/2.81/\textbf{86.01}/23.2$^\dagger$\\
21 &\cmark  &  &  & \multirowcell{1}{\scriptsize{ECAPA-TDNN}}  &   5.07/2.52/77.57/44.7 & \textbf{9.82/2.91}/86.06/\textbf{28.8} & 10.95/\textbf{2.98/88.24}/25.9 & \textbf{11.26/3.07/89.14}/24.5 & \textbf{9.25/2.87}/85.22/\textbf{31.1}$^\dagger$  & \textbf{8.30/2.82}/85.96/\textbf{22.5}$^\dagger$\\
\midrule[1.2pt]
\end{tabular}
}
\vspace{-0.7cm}
\end{table*}

\label{sec:Experimental_Results}
\begin{figure}[htbp]
    \centering
    \setlength{\abovecaptionskip}{0pt plus 1pt minus 3pt}
    \includegraphics[scale=0.7]{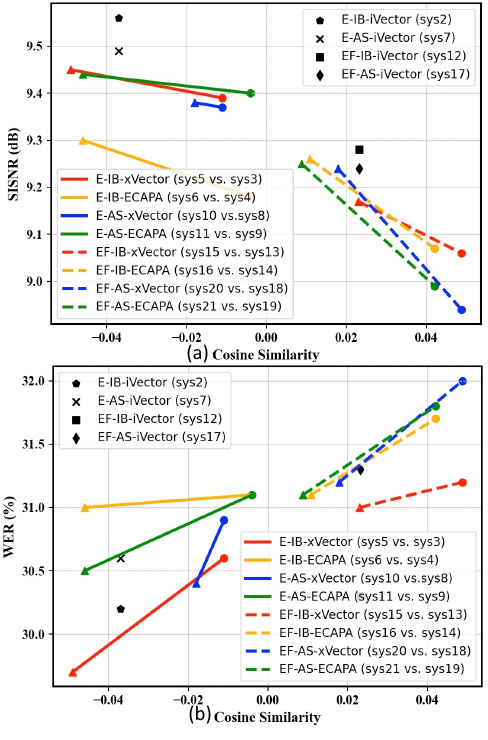}
    \caption{Correlation between cosine similarity and SISNR or overall WER on "Dev" set for systems in Table~\ref{tab:table2}.  Comparable systems with/without joint and non-joint speaker features learning marked as ``$\triangle$"  and ``$\bigcirc$" respectively at two ends of each colored line. ``EF", ``E", ``IB", and ``AS" denote ``enrollment free", ``enrollment", ``input bias" and ``activation scaling".
    }
    \label{fig:cosine_simi}
 \vspace{-0.7cm}
\end{figure}

\section{Experiments}
\vspace{-0.1cm}
\subsection{Experimental Setup}
\vspace{-0.1cm}
\noindent\textbf{Simulated Multichannel Mixture Speech:} 
Experiments were conducted on the overlapped and noisy speech simulated using the LRS3-TED dataset \cite{afouras2018lrs3}. 
A 15-channel symmetric linear array with non-even inter-channel spacing [7,6,5,4,3,2,1,1,2,3,4,5,6,7]cm is used in the simulation process.
843 point-source noises \cite{ko2017study} and 20000 room impulse responses (RIRs) generated by the image method \cite{habets2006room} in 400 different simulated rooms are used in our experiment. 
The distance between a sound source and the microphone array center is uniformly sampled from a range of 1m to 5m and the room size ranges from 4m$\times$4m$\times$3m to 10m$\times$10m$\times$6m (length$\times$width$\times$height).
The reverberation time $T_{60}$ is uniformly sampled from a range of 0.14s to 0.92s. 
The average overlapping ratio is around 80\%.
The signal-to-noise ratio (SNR) is uniformly sampled from \{0, 5, 10, 15, 20\}dB, and the signal-to-interference ratio (SIR) is uniformly sampled from \{-6, 0, 6\}dB. 
In addition, the angle difference relative to the microphone array between the target and interfering speakers is uniformly sampled from four ranges of the angle difference
\{[$0^{\circ}$, $15^{\circ}$), [$15^{\circ}$, $45^{\circ}$), [$45^{\circ}$, $90^{\circ}$), [$90^{\circ}$, $180^{\circ}$)\}. 
The final simulated multichannel datasets contain four subsets with 110354, 3122, 3136 and 1321 utterances each for training (111.47 hours, 4886 speakers), val (3.15 hours, 157 speakers),  dev (3.14 hours, 158 speakers) and test (0.84 hours, 412 speakers). 
\\
\noindent\textbf{Baseline System Description:}
The configuration settings of the baseline speech separation front-end, visual feature extraction and Conformer ASR back-end are referred to \cite{li2023audio}.\\
\noindent\textbf{Implementation Details:} 
Speaker features are empirically configured for iVector, xVector or ECAPA-TDNN as follows: \textbf{a)} speaker features are fed into the system at the input and output of TCN0 Audio Block (Fig. 1, upper middle, in pink) respectively for input bias and activation scaling based fusion; \textbf{b)} features dimensionality set as 100, 256 and 256 respectively for iVector, xVector and ECAPA-TDNN in all the experiments.

\vspace{-0.3cm}
\subsection{Experimental Results}
\vspace{-0.2cm}
\noindent\textbf{Performance of joint speaker features learning} is shown in Table \ref{tab:table2}. Several trends can be found. 
\textbf{1)} The proposed joint speaker features learning brings improvements consistently for both enrollment-based and enrollment-free zero-shot adaptation compared to non-joint speaker features learning (sys. 5,6,10,11,15,16,20,21 vs. sys. 3,4,8,9,13,14,18,19), with a statistically significant WER reduction up to \textbf{0.7\%} and \textbf{1.1\%} absolute (\textbf{2.2\%} and \textbf{4.7\%} relative) (sys. 21 vs. sys. 19) on the dev and test sets. Consistent improvements are also found on speech enhancement metrics (SISNR~\cite{bahmaninezhad2019}, PESQ~\cite{recommendation2001perceptual} and STOI~\cite{taal2011algorithm}) scores. 
\textbf{2)} These improvements are consistently correlated with inter-speaker discrimination measured in cosine similarity.  
In terms of the correlation between cosine similarity and SISNR in Fig \ref{fig:cosine_simi}(a) or overall WER in Fig \ref{fig:cosine_simi}(b),  using joint speaker features learning can boost the system performance characterized by lower cosine similarity scores\footnote{Cosine similarity score is computed using the speaker features extracted from the target and interfering speaker's speech respectively.}.
\textbf{3)} The performance of enrollment-free adaptation is comparable to more expensive enrollment-based adaptation when varying the inter-speaker angle difference between $15^{\circ}$ to $180^{\circ}$, except for the most challenging subset when the inter-speaker angle difference is under $15^{\circ}$. 
This is expected as the quality of SI system speech separation outputs, which are used to extract target speaker features, is heavily degraded and thus negatively impacts speaker adaptation performance. How to mitigate the sensitivity to inter-speaker angle difference for enrollment-free adaptation will be studied in future research.\\
\noindent\textbf{Performance of speaker adaptation for end-to-end speech separation and recognition} is shown in Table \ref{tab:table3}. 
The baseline system (sys. 1, in Table \ref{tab:table2}) and the respective best-performed systems from both enrollment-based and zero-shot enrollment-free adaptation (sys. 2,21, in Table \ref{tab:table2}) are used correspondingly for end-to-end audio-only systems training (sys. 1,2,3, in Table \ref{tab:table3}).  Such selected systems are jointly fine-tuned except for using a stronger Conformer ASR back-end integrated with fine-tuned WavLM SSL features. 
When applying speaker adaptation to both stages, the proposed speaker-adaptive audio-visual systems consistently outperformed the SI baseline by statistically significant WER reductions of up to \textbf{2.3}\% and \textbf{2.7}\% absolute (\textbf{18.1}\% and \textbf{35.1}\% relative) on dev and test sets (sys. 5 vs. sys. 4).  
It can be observed that the proposed best-performed speaker-adaptive audio-visual system (sys. 5) can significantly outperform the baseline fine-tuned WavLM Model using the raw first channel of overlapped speech by statistically significant WER reductions of up to \textbf{21.6\%} and \textbf{25.3\%} absolute (\textbf{67.5\%} and \textbf{83.5\%} relative) across dev and test sets.
\vspace{-0.4cm}

\begin{table}[h!]
\centering
\caption{Performance of speaker-adaptive audio-visual end-to-end speech separation and recognition systems integrated with fine-tuned WavLM SSL features.  ``Sep."  and ``Recog." denote ``separation" and ``recognition". ``$\ast$", $\dagger$ and ``$\ddag$"  represent a statistically significant WER difference (MAPSSWE, $\alpha$=0.05\cite{Gillick1989SomeSI} ) over the audio-only SI baseline (sys. 1), audio-visual SI baselines (sys. 4) and baseline fine-tuned WavLM model.
}
\vspace{-0.35cm}
\label{tab:table3}
\resizebox{0.89\columnwidth}{!}{
\begin{tabular}{c|c|c|c|c|c|c|c} 
 \toprule[1pt]
 \multirowcell{2}{Sys.} &  
 \multirowcell{2}{\tabincell{c}{Adaptation \\ Supervision}} &
  \multicolumn{2}{c|}{+Visual}  & 
 \multicolumn{2}{c|}{+Speaker Adaptation} & 
 \multicolumn{2}{c}{WER $(\downarrow)$}  \\ 
\cline{3-8}
 &   & \multirowcell{1}{Sep.}&  \multirowcell{1}{Recog.} & \multirowcell{1}{Sep.}  & \multirowcell{1}{Recog.}   & \multirowcell{1}{Dev} & \multirowcell{1}{Test}  \\
\midrule[1pt]
\multicolumn{6}{c|}{Fine-tuned WavLM Model by raw first channel of overlapped speech} & 32.0 & 30.3 \\
\hline
1 & SI Baseline & \multirowcell{3}{\xmark} & \multirowcell{3}{\xmark}  &  \xmark & \xmark & 15.1 & 11.2\\
2 & Enroll.    & & & \cmark & \cmark &     12.1$^{\ast \ddag}$& 7.9$^{\ast \ddag}$\\
3 & Enroll. Free & & & \cmark & \cmark & 14.1$^{\ast \ddag}$ & 10.1$^{\ast  \ddag}$\\ \hline
4 & SI Baseline & \multirowcell{3}{\cmark} & \multirowcell{3}{\cmark} & \xmark & \xmark & 12.7  & 7.7\\
5 & Enroll. &  &  &  \cmark & \cmark  & 10.4$^{\dagger \ddag}$ & 5.0$^{\dagger \ddag}$  \\
6 & Enroll. Free&  &  & \cmark & \cmark  & 12.5 $^\ddag$ & 7.0 $^\ddag$  \\

\bottomrule[1pt]
\end{tabular}
}
\vspace{-0.5cm}
\end{table}

\vspace{-0.3cm}
\section{Conclusions}
\vspace{-0.2cm}
This paper proposes joint speaker feature learning methods for zero-shot adaptation of audio-visual multichannel speech separation and recognition systems. xVector and ECAPA-TDNN speaker encoders are connected using purpose-built fusion blocks and tightly integrated with the complete system training. Experiments conducted on LRS3-TED data simulated multichannel overlapped speech suggest that joint speaker feature learning consistently improves speech separation and recognition performance over the baselines without using such.
Further analyses reveal performance improvements are strongly correlated with increased inter-speaker discrimination measured using cosine similarity. 
Future research will focus on mitigating the sensitivity to inter-speaker angle differences for enrollment-free adaptation.

\section{Acknowledgements}
This research is supported by Hong Kong RGC GRF grant No. 14200021, 14200220, TRS under Grant T45-407/19N, Innovation \& Technology Fund grant No. ITS/218/21, National Natural Science Foundation
of China (NSFC) Grant 62106255, and Youth Innovation Promotion Association CAS Grant 2023119.


\bibliographystyle{IEEEtran}
\bibliography{mybib}

\end{document}